\documentclass[aps,prc,twocolumn,superscriptaddress,showpacs,floatfix,nofootinbib]{revtex4-1}
\usepackage{amsmath,graphicx,color}
\newcommand{\beq}{\begin{equation}}
\newcommand{\eeq}{\end{equation}}
\newcommand{\bea}{\begin{eqnarray}}
\newcommand{\eea}{\end{eqnarray}}

\begin{document}

\title{Viscous corrections to anisotropic flow and transverse momentum
spectra from transport theory}

\author{Salvatore Plumari}
\author{Giovanni Luca Guardo}
\author{Vincenzo Greco}
\affiliation{Department of Physics and Astronomy, University of Catania,
Via S. Sofia 64, I-95125 Catania}
\affiliation{Laboratorio Nazionale del Sud, INFN-LNS, Via S. Sofia 63,
I-95125 Catania}
\author{Jean-Yves Ollitrault}
\affiliation{
Institut de physique th\'eorique, Universit\'e Paris Saclay, CNRS, CEA, F-91191 Gif-sur-Yvette, France} 
\date{\today}

\begin{abstract}
Viscous hydrodynamics is commonly used to model the
evolution  of the matter created in an ultra-relativistic heavy-ion collision. 
It provides a good description of transverse momentum 
spectra and anisotropic flow. 
These observables, however, cannot be consistently derived using 
viscous hydrodynamics alone, because they depend on the microscopic
interactions at freeze-out.  
We derive the ideal hydrodynamic limit and the first-order viscous
correction to anisotropic flow ($v_2$, $v_3$ and 
$v_4$) and momentum spectrum using a transport calculation. 
The linear response coefficient to the initial anisotropy, 
$v_n(p_T)/\varepsilon_n$, depends little on $n$ in the ideal
hydrodynamic limit. 
The viscous correction to the spectrum depends not only on
the differential cross section, but also on the initial momentum
distribution. This 
dependence is not captured by standard second-order viscous hydrodynamics. 
The viscous correction to anisotropic flow increases with $p_T$, but
this increase is slower than usually assumed in viscous hydrodynamic
calculations.  In particular, 
it is too slow to explain the observed maximum of $v_n$ at 
$p_T\sim 3$~GeV/c. 
\end{abstract}

\maketitle

\section{Introduction}
Relativistic viscous hydrodynamics~\cite{Heinz:2013th,Gale:2013da} is
the state of the art for describing the evolution of the
strongly-coupled system formed in an ultrarelativistic heavy-ion
collision at RHIC or LHC.   
It has long been realized \cite{Kolb:2003dz} 
that ideal hydrodynamics naturally explains the large magnitude of 
elliptic flow~\cite{Adler:2003kt,Aamodt:2010pa}.  
However, the system formed in such a collision is so small that
deviations from local thermal equilibrium are sizable, resulting in
the inclusion of viscosity~\cite{Romatschke:2007mq} in hydrodynamic
calculations. Viscosity typically reduces the magnitude of elliptic
flow by 20\%~\cite{Luzum:2008cw}. 
Viscous effects on higher harmonics of anisotropic
flow~\cite{ALICE:2011ab,Adare:2011tg}, such as triangular
flow~\cite{Alver:2010gr}, are even larger~\cite{Alver:2010dn}. 

Even though there is a consensus that viscosity matters, the
calculation of viscous corrections to observables is not yet under
control. The reason is that viscosity affects not only the space-time
history of the fluid~\cite{Baier:2007ix}, but also the momentum
distribution of particles at ``freeze-out'', which has an 
off-equilibrium part proportional to viscosity~\cite{Teaney:2003kp,Teaney:2013gca}. 
Viscous hydrodynamics itself does not fully specify this
off-equilibrium part. 
The only requirement is that the system of particles should generate
the same energy-momentum tensor as the fluid just before freeze
out~\cite{Luzum:2010ad}.  
This requirement, however, does not constrain the dependence of 
the relative deviation to equilibrium on the momentum $p$ in the
rest frame of the fluid, which is essentially a free function $\chi(p)$. 
This function is not universal, and involves 
the differential cross sections between constituents~\cite{Dusling:2009df}. 
It is typically put by hand in hydrodynamic calculations. 

The common lore is that effects of viscosity are more important for
particles with larger transverse momenta~\cite{Gale:2013da}. 
This is due to the fact that most hydrodynamic calculations use the
``quadratic'' ansatz~\cite{Teaney:2003kp} $\chi(p)\propto p^2$. 
While this choice generally results in a improved description of
experimental data~\cite{Gale:2013da}
(see however~\cite{Luzum:2010ad}), it is not supported by any
theoretical argument~\cite{Dusling:2009df}. 
We evaluate viscous corrections to observables 
(specifically, transverse momentum spectra and anisotropic flow)
by solving numerically a relativistic Boltzmann equation. 
We simulate relativistic particles
undergoing $2\to 2$ elastic collisions with a total cross section
$\sigma_{\rm tot}$. 
In the limit $\sigma_{\rm tot}\to +\infty$, a generic observable $f(\sigma)$
can be expanded in powers of $1/\sigma_{\rm tot}$:
\begin{equation}
\label{asymptotic}
f(\sigma_{\rm tot})\underset{\sigma_{\rm tot}\to\infty}{\approx}
f^{(0)}+\frac{1}{\sigma_{\rm tot}}\delta f+
{\cal O}\left(\frac{1}{\sigma_{\rm tot}^2}\right).
\end{equation}
The leading term $f^{(0)}$ is the limit of infinite cross section, which
corresponds to ideal hydrodynamics in the limit of a vanishing
freeze-out temperature~\cite{Borghini:2005kd,Gombeaud:2007ub}.
The next-to-leading term $\delta f$ is a viscous correction: since 
the shear viscosity $\eta$ scales like $1/\sigma_{\rm tot}$~\cite{Plumari:2012ep},
this correction is proportional to $\eta$.\footnote{Note that in
  hydrodynamic calculations, $\delta f$ often denotes the viscous
  correction at freeze-out. Here, $\delta f$ is the full viscous
  correction, which also contains a contribution from the hydrodynamic
  evolution.}

We evaluate $f^{(0)}$ and $\delta f$ by solving
numerically the relativistic Boltzmann equation for several large 
values of the cross section $\sigma_{\rm tot}$. 
Our primary goal is to illustrate by an explicit calculation how the viscous 
correction to anisotropic flow depends on transverse momentum $p_T$,
and to what extent this dependence is sensitive to the structure of
the differential cross section. 
We do not mean here to carry out a full realistic simulation of a
heavy-ion collision. In particular, for sake of simplicity, our transport
calculation uses massless particles which supply the possibility
of having only shear viscosity with no bulk viscosity 
~\cite{Monnai:2009ad,Bozek:2009dw,Noronha-Hostler:2013gga}.
The resulting equation of state is harder~\cite{Ollitrault:2008zz}
than that of QCD near the deconfinement crossover~\cite{Borsanyi:2013bia}. 
This results in larger $v_n$ and harder $p_T$ spectra. 

We also study the dependence of observables on initial conditions. 
In second-order viscous hydrodynamics, the evolution
is completely specified by the initial value of the 
energy-momentum tensor $T^{\mu\nu}$~\cite{Baier:2007ix,Gale:2013da}.  
At the microscopic level, however, the initial momentum distribution 
contains more information than just $T^{\mu\nu}$. Usual viscous
hydrodynamics assumes that 
this additional information is washed out by the system evolution. Our
simulation provides a means of testing this assumption, by 
constructing two different initial conditions with the exact same
$T^{\mu\nu}$, and comparing the observables at the end of the
evolution. 

\section{Initial conditions and evolution}

Initial conditions follow Bjorken's boost-invariant
prescription~\cite{Bjorken:1982qr}, but with a finite extent in
space-time rapidity $-2.5<\eta<2.5$. 
The evolution is started  
at time $\tau_0 = 0.6$~fm/$c$~\cite{Heinz:2001xi} after the collision. 
The initial conditions of the Boltzmann equation  are specified by the
one-body density $f(x,p)$ in coordinate ($x$) and momentum
($p$) space at time $\tau_0$. 

The initial density profile in transverse coordinate space $(x,y)$ is
taken from an optical Glauber~\cite{Miller:2007ri} calculation for a
central Au-Au collision at $\sqrt{s}=200$~GeV, corresponding to the
top RHIC energy.
This initial density is azimuthally symmetric, so that 
anisotropic flow vanishes by construction. 
We introduce anisotropy artifically by deforming the initial
distribution, thus mimicking an initial state
fluctuation~\cite{Alver:2010gr}. 
In hydrodynamics, one typically deforms the initial energy 
density profile~\cite{Alver:2010dn,Retinskaya:2013gca}. 
In a transport calculation, where the initial conditions are specified
by the initial positions of particles, it is simpler to just shift
these positions by a small amount. Introducing the complex notation
$z=x+iy$, in order to generate flow in
harmonic $n$, we shift $z$ according to 
\begin{equation}
\label{deformation}
z\to z+\delta z\equiv z-\alpha \bar z^{n-1},
\end{equation}
where 
$\bar z\equiv x-iy$, and $\alpha$ is a real positive quantity chosen in
such a way that the correction is small. 
This transformation is invariant under the change 
$z\to e^{2i\pi/n}z$, i.e., it has $2\pi/n$ symmetry. 
Therefore, to leading order in $\alpha$, the only nonvanishing
anisotropic flow coefficient is $v_n$.\footnote{Note that the
deformation used in 
Refs.~\cite{Alver:2010dn,Retinskaya:2013gca} is singular at the origin
$z=0$ for $n\ne 2$. By contrast, the deformation defined by
Eq.~(\ref{deformation}) is regular at the origin, but has a
singularity at large $|z|$ (see also~\cite{Teaney:2010vd}). On the $x$
axis, the singularity is 
located  at the point where $\partial \delta x/\partial x=-1$. 
If $\alpha$ is small, this singularity occurs at a point where
the density is low.}
The initial eccentricity in harmonic $n$ is defined for $n\ge 2$
by~\cite{Teaney:2010vd,Bhalerao:2011yg} 
\begin{equation}
\label{defepsn}
\varepsilon_n\equiv -\frac{\sum_j (z_j+\delta z_j)^n}{\sum_j
|z_j+\delta z_j|^n},
\end{equation}
where the sum runs over all particles with initial position $z'_j$. 
Inserting Eq.~(\ref{deformation}) into Eq.~(\ref{defepsn}), and using the
fact that the distribution of $z_j$ is 
azimuthally symmetric, one obtains, to leading order in $\alpha$ and
for a large number of particles 
\begin{eqnarray}
\varepsilon_n&\simeq& -\frac{\sum_j n z_j^{n-1}\delta z_j}{\sum_j
|z_j|^n}
=n\alpha\frac{\sum_j |z_j|^{2(n-1)}}{\sum_j |z_j|^n}\cr
&\simeq &n\alpha\frac{\langle r^{2(n-1)}\rangle}{\langle r^n\rangle},
\end{eqnarray}
where, in the right-hand side, angular brackets denote an average
taken with the optical Glauber profile, and $r=|z|=\sqrt{x^2+y^2}$. 
We carry out simulations for $n=2,3,4$. For each $n$, we fix
$\alpha$ in such a way that $\varepsilon_n=0.2$. 
We have checked that this value is sufficiently small that the
response is linear~\cite{Alver:2010dn}.
In hydrodynamics with fluctuating initial conditions, $v_2$ and $v_3$
are determined to a good approximation by linear response to the
eccentricity in the corresponding
harmonic~\cite{Niemi:2012aj,Gardim:2014tya}.  
On the other hand, $v_4$ is the superposition of a linear
term~\cite{Teaney:2012ke}, and a nonlinear term induced by
$v_2$~\cite{Gardim:2011xv}. The present study only addresses the
linear term.

In momentum space, we consider two different types of initial
conditions. 
The first case is a thermal Boltzmann distribution 
\begin{equation}
\label{thermal}
dN/d^3p\propto \exp(-p/T). 
\end{equation}
In addition, one requires that the temperature $T$ and the local
density $n$ be such that $n/T^3$ is a constant throughout the
transverse plane, as in a thermal gas of massless particles with zero 
chemical potential. Thus this initial condition is that of a  
hydrodynamic calculation with zero chemical potential~\cite{Ollitrault:2008zz}. 
The initial temperature at the center of the fireball is $T_0 =
340$~MeV. 
The second case is a constant distribution:
\begin{equation}
\label{constant}
dN/d^3p\propto \theta(p_0-p). 
\end{equation}
The maximum momentum $p_0$ and the
proportionality constant are chosen such that the particle density $n$
and the energy density $\varepsilon$ are the same as with the previous
initial condition. 
Similar initial conditions have been previously used in transport
calculations in order to mimic the effect of saturation in high-energy
QCD~\cite{Blaizot:2011xf,Epelbaum:2014mfa}. 
Both types of initial conditions have exactly the same energy-momentum
tensor. 

The evolution of the system is determined by 
the relativistic classical Boltzmann equation. 
We use a relativistic transport code developed to study  heavy-ion 
collisions at RHIC and LHC
energies
\cite{Ferini:2008he,Greco:2008fs,Plumari:2011re,Plumari:2012xz,Plumari:2012ep,Ruggieri:2013bda},  
which uses the test-particle method. The collision integral is
solved by using Monte Carlo methods based on the stochastic
interpretation of transition
amplitudes \cite{Xu:2004mz,Ferini:2008he,Plumari:2012ep}.  
The total cross section is fixed throughout the evolution. 
The differential cross section is 
\begin{equation}
 \frac{d\sigma}{dt} \propto \frac{1}{\left(t-m_D^2\right) ^2},
\label{sigma_md}
\end{equation}
where $t$ is the usual Mandelstam variable. 
This differential cross section is typically used in parton cascade
approaches  
\cite{Zhang:1999rs,Molnar:2001ux,Ferini:2008he,Greco:2008fs,Xu:2004mz}
and by symmetry the $u$-channel is included. 
The limit $m_D\to\infty$ corresponds to an isotropic cross
section. The opposite limit, where  $m_D$ is smaller than the typical
particle energy, corresponds to a forward-peaked cross section. 

In a transport approach, one can follow the evolution 
of the system until the last collision, but this is numerically
expensive. Instead, we choose to follow the system until a fixed final time, 
and check stability of our results with respect to this final time. 
The calculations presented in this paper are carried out with a final
time $t_f=8$~fm/$c$, but we have checked that the momentum spectra are
unchanged if we extend the final time to $t_f=12$~fm/$c$. 
Since the initial density profile possesses $\phi\to -\phi$ symmetry 
(where $\phi$ is the azimuthal angle), 
we define anisotropic flow as $v_n=\langle\cos n\phi\rangle$, 
where angular brackets denote an average over particles at the end of
the evolution. 

Throughout this paper, we carry out four sets of calculations: 
Three sets with the thermal initial distribution (\ref{thermal}) and
the values $m_D=0.3$~GeV, $m_D=0.7$~GeV and an isotropic cross
section, and a fourth set with the constant initial distribution
(\ref{constant}) and an isotropic cross section. 
Thus we study how results depend on the initial distribution
and on the differential cross section.

\begin{figure}
\begin{center}
\includegraphics[width=\linewidth]{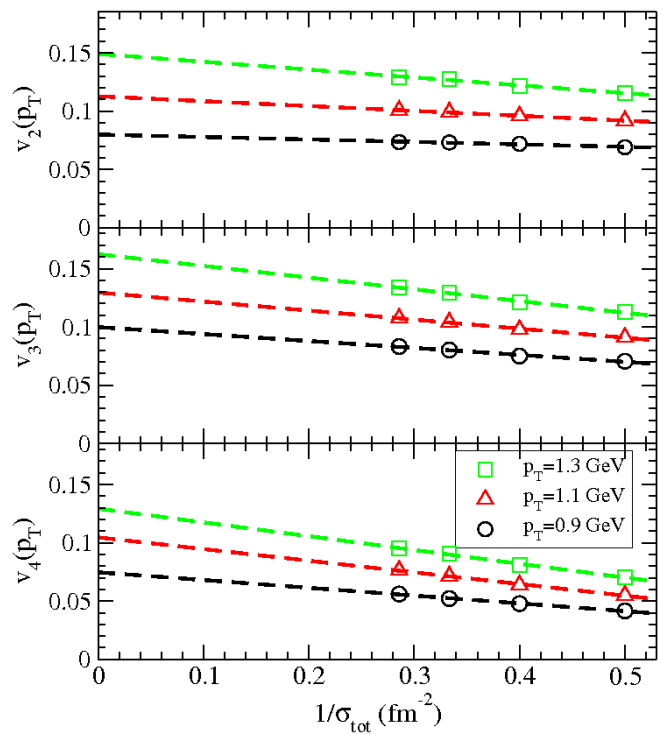}
\end{center}
\caption{Left panel: $v_n(p_T)$ as a function of $1/\sigma_{\rm tot}$ (thermal
  initial distribution, isotropic scattering cross section) in the
  rapidity interval $|y|<0.5$. The
  $y$-intercept is the ideal hydrodynamic limit $v_n^{(0)}$, while the
  slope corresponds to the viscous correction $\delta v_n$, as defined
  by Eq.~(\ref{asymptotic}).}
\label{Fig:extrapolation}
\end{figure}
For each set of parameters, we perform different calculations for the
following set of total cross sections: $\sigma_{\rm tot}= 20$, 25, 30 and 35~mb. 
Results for anisotropic flow $v_n$~\cite{Heinz:2013th} are shown in
Fig.~\ref{Fig:extrapolation}. 
The dependence of these observables on $1/\sigma_{\rm tot}$ is essentially
linear, corresponding to the regime where 
viscous hydrodynamics applies. In order to improve the accuracy, we 
fit these results with a polynomial of order 2 in $1/\sigma_{\rm tot}$
and  extract the ideal hydrodynamic limit and the first
viscous correction using Eq.~(\ref{asymptotic}):
specifically, the ideal hydrodynamic limit $f^{(0)}$ is the
extrapolation to $1/\sigma_{\rm tot}\to 0$ and the viscous correction $\delta f$
is the slope at the origin.

\section{Ideal hydrodynamics}

\begin{figure}
\begin{center}
\includegraphics[width=\linewidth]{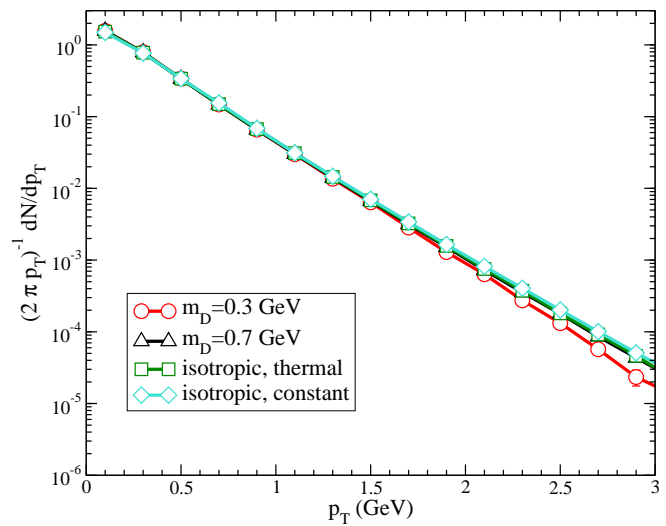}
\end{center}
\caption{Ideal hydrodynamic limit for the $p_T$ spectrum 
in the rapidity interval
  $|y|<0.5$. The spectra have been divided by $dN/dy$ so that they are
  normalized to unity.}
\label{Fig:f0}
\end{figure}
We first study the ideal hydrodynamic limit, defined by
the extrapolation $\sigma_{\rm tot}\to+\infty$ in the Boltzmann equation. 
Ideal hydrodynamics corresponds to local thermal 
equilibrium~\cite{Ollitrault:2008zz}. 
One expects thermalization to wash out details of initial conditions,
so that observables should not depend on the initial momentum
spectrum provided the energy density is fixed. 
Similarly, the momentum distribution in thermal equilibrium is
universal, therefore one expects observables to be independent of the
differential cross section in this limit. 

Figure~\ref{Fig:f0} displays the ideal hydrodynamic limit for the
transverse momentum distribution in the rapidity interval $|y|<0.5$. 
As expected, it is independent of initial conditions and of 
the differential cross section. 
The distribution is essentially exponential in $p_T$, with a slight 
upward curvature which is a typical consequence of collective
motion~\cite{Kolb:2003dz,Retiere:2003kf}. 
\begin{figure}
\begin{center}
\includegraphics[width=\linewidth]{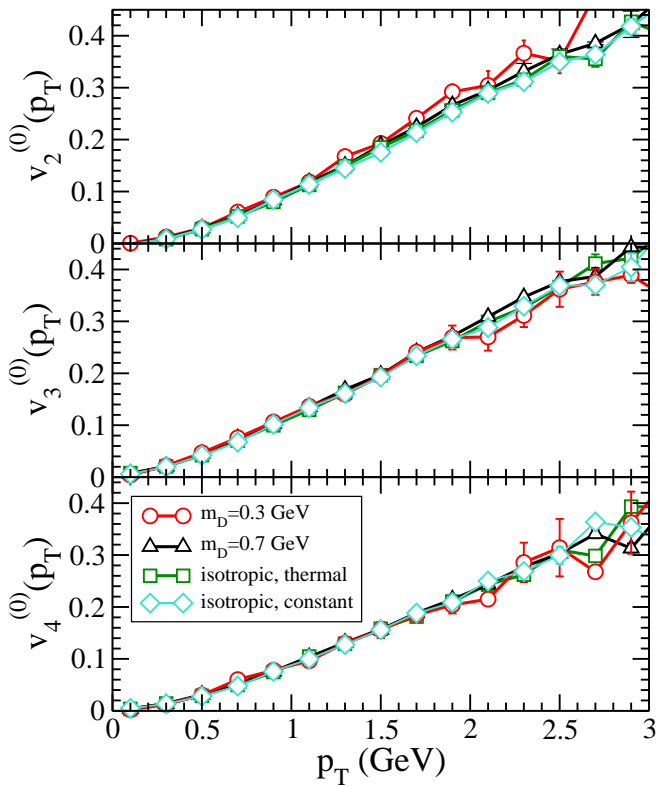}
\end{center}
\caption{Ideal hydrodynamic limit of anisotropic flow $v_n$.
Top to bottom: $n=2,3,4$. For all harmonics, we have chosen
$\varepsilon_n=0.2$, so that the linear response coefficients
$\kappa_n\equiv v_n/\varepsilon_n$ are larger by a
factor 5.} 
\label{Fig:vn0}
\end{figure}
The ideal hydrodynamic limit of anisotropic flow coefficients is
displayed in Fig.~\ref{Fig:vn0}.  
$v_n(p_T)$ is linear at high $p_T$, as observed in previous
numerical calculations~\cite{Alver:2010dn}.  
This behavior is general in ideal hydrodynamics~\cite{Teaney:2012ke}
in the linear regime $\varepsilon_n\ll 1$.  
Interestingly, the magnitude of the linear response
$\kappa_n(p_T)\equiv v_n(p_T)/\varepsilon_n$~\cite{Yan:2014nsa}  is approximately 
the same for $n=2,3,4$. 
It would be interesting to reproduce these results using ideal
hydrodynamics with the same ideal gas equation of state
(constant sound velocity $c_s=1/\sqrt{3}$) and a small freeze-out
temperature.

\section{Viscous correction}

\begin{figure}
\begin{center}
\includegraphics[width=\linewidth]{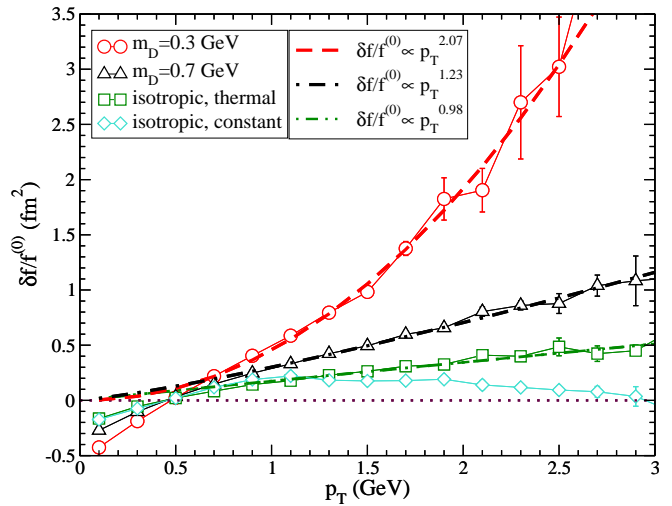}
\end{center}
\caption{First-order relative viscous correction to the normalized
  transverse momentum distribution in the rapidity interval
  $|y|<0.5$. 
Dashed lines are power-law fits. }
\label{Fig:df_f}
\end{figure}
We now present results for the first-order viscous correction to
observables, corresponding to the term $\delta f$ in
Eq.~(\ref{asymptotic}).  
We scale the viscous correction $\delta f$ by the ideal hydrodynamic
limit $f^{(0)}$. 
Eq.~(\ref{asymptotic}) shows that $\delta f/f^{(0)}$ has the dimension
of the cross section $\sigma_{\rm tot}$. 
The relative viscous correction is $\delta f/f^{(0)}$ divided by the
total cross section $\sigma_{\rm tot}$.
As a rule of thumb, viscous hydrodynamic applies if 
$\delta f/f^{(0)}$ is significantly smaller than $\sigma_{\rm tot}$
in absolute value.

Figure~\ref{Fig:df_f} displays  $\delta f/f^{(0)}$ for the transverse
momentum spectrum. 
Viscous effects result in a particle excess at large $p_T$, 
corresponding to an increase of the average $p_T$, that is, a higher
effective temperature. 
The reason is that viscosity decreases the longitudinal pressure, thereby
reducing longitudinal cooling\cite{Chaudhuri:2005ea,Florkowski:2010cf}. 

Unlike the ideal hydrodynamic limit, the first-order viscous
correction depends on the differential cross section. 
It is larger for smaller values of $m_D$. This is 
due to the fact that the scattering is forward peaked and less 
efficient in thermalizing the system. 
For an isotropic cross section, the relative viscous correction is
almost linear in $p_T$: 
$\delta f/f^{(0)}\propto {p_T}^{0.98}$. This result is consistent with
the results obtained in \cite{Bhalerao:2013pza} in the Chapman-Enskog
approximation. 

Surprisingly, the viscous correction also depends on the initial 
momentum distribution. Two initial conditions with exactly the same
energy-momentum tensor $T^{\mu\nu}$ lead to different first-order viscous
corrections to observables, at variance with usual viscous
hydrodynamics~\cite{Baier:2007ix}.  
The reason is that the constant momentum distribution (\ref{constant})
is strongly out of equilibrium, therefore hydrodynamics does not apply
at early times. 
Figure~\ref{Fig:df_f} shows that with this constant distribution, 
$\delta f$ is smaller at high $p_T$ than with a thermal initial
distribution. 
This depletion at high $p_T$ can be understood as a memory of
the initial conditions, where all particles initially have momenta
below a threshold $p_0$.  

\begin{figure}
\begin{center}
\includegraphics[width=\linewidth]{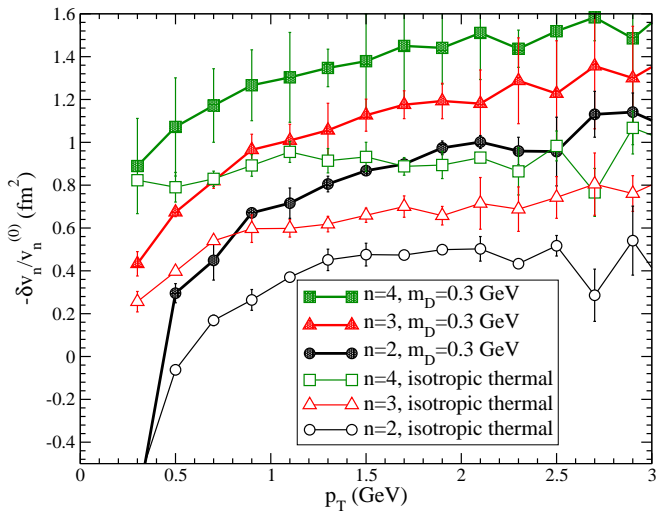}
\end{center}
\caption{$-\delta v_{n}(p_T)/v_n^{(0)}$ as a function of transverse
  momentum $p_T$ in the rapidity interval $|y|<0.5$. 
Open symbols refer to the case of isotropic cross
section and full symbols to a forward-peaked
cross section. The circles, triangles and squares
refer to n = 2, 3 and 4 respectively.} 
\label{Fig:dvn}
\end{figure}

Finally, we study the viscous correction to anisotropic flow
$v_n(p_T)$. 
Figure~\ref{Fig:dvn} displays $-\delta v_{n}/v_n^{(0)}$
as a function of transverse momentum $p_T$.
Viscous effects decrease anisotropic flow~\cite{Teaney:2003kp}, 
therefore the correction is shown with a minus sign. 
The viscous correction to $v_n$ increases as a function of harmonic
order $n$, as already observed in viscous hydrodynamic
calculations~\cite{Alver:2010dn}. 
At large transverse momentum, it scales approximately like 
the order $n$~\cite{Hatta:2014jva}, $-\delta
v_{n}(p_T)/v_n^{(0)}\propto n$. 
This dependence is weaker than the $n^2$ dependence reported in
previous studies~\cite{Gubser:2010ui,Staig:2010pn,Lacey:2013qua,Teaney:2012ke}.  

As expected, the viscous correction depends on the differential cross
section. As for the spectrum, it is larger with a forward-peaked cross
section ($m_D=0.3$~GeV) than with an isotropic cross section. 
Results with the thermal distribution (\ref{thermal}) and with the
constant distribution (\ref{constant}) (not shown) are 
consistent within error bars, thereby suggesting that the viscous
correction to $v_n$ is roughly independent of the initial momentum
distribution. 

With an isotropic cross section, the relative viscous correction saturates
at high $p_T$, 
while the usual quadratic freeze-out ansatz~\cite{Teaney:2003kp}
predicts a linear increase. 
A mild increase with $p_T$ is observed with a forward-peaked cross
section, but it is still much slower than linear. 
Note that our viscous correction must be compared with the full
viscous correction in hydrodynamics, which results in part from the
hydrodynamic evolution, and in part from the freeze-out. With
a quadratic freeze-out ansatz, however, the behavior of $v_n$ at large
$p_T$ in hydrodynamics is dominated by the correction at freeze-out. 
Thus the quadratic freeze-out ansatz currently used in most viscous 
hydrodynamic calculations is not supported by microscopic transport
calculations. 

\section{Conclusions}
We have calculated the ideal hydrodynamic limit and the first-order
viscous correction for the transverse momentum ($p_T$) distribution
transverse-momentum spectra and anisotropic flow $v_n(p_T)$ 
by solving numerically a Boltzmann equation and
studying the limit of large scattering cross section $\sigma_{\rm tot}\to\infty$. 
The ideal hydrodynamic limit is found to be independent of
microscopic details, as expected from the universality of
thermodynamic behavior. 
The linear response coefficients $v_n(p_T)/\varepsilon_n$ depends
little on harmonic $n$ in the ideal hydrodynamic limit. 

The first order viscous corrections to observables, on the other hand,
are not universal. 
As expected~\cite{Dusling:2009df}, they depend on the differential
cross section. 
For all the differential cross sections investigated in this paper, 
we find that the relative viscous correction to
anisotropic flow, $v_n$, does not increase significantly with $p_T$ at
large $p_T$. Our results suggest that first-order viscous
corrections do {\it not\/} explain the decrease of $v_n$ at high $p_T$, at
variance with common lore \cite{Gale:2013da}, and that a different
mechanism, such as jet quenching~\cite{Gyulassy:2003mc}, is needed at high
$p_T$. 

The viscous correction to $v_n$ increases linearly 
with $n$. The stronger ($n^2$) dependence typically found in
hydrodynamics~\cite{Teaney:2012ke}  leads to negative values of $v_4$
and $v_5$ at large $p_T$~\cite{Teaney:2013gca}, even for small
viscosities. A weaker dependence on harmonic $n$ is therefore likely
to improve agreement with experimental data. 

Finally, our results clearly show that usual, second-order
relativistic hydrodynamics does not fully 
capture the first-order viscous correction to observables. Two
initial conditions with exactly the same energy-momentum tensor, which
would therefore yield the exact same hydrodynamical flow, are found to
yield different momentum spectra at the end of the evolution. 
Specifically, the first-order viscous correction is found to retain
the memory of the initial condition. 
Within a strong coupling calculation, a proper treatment of the underlying 
microscopic degrees of freedom leads to an evolution equation which is 
second-order rather than first-order~\cite{Heller:2014wfa}, so that
the solution is not solely determined by the initial value of $T^{\mu\nu}$. 
Our calculation provides an explicit illustration, within 
a weak-coupling calculation, that the initial value of $T^{\mu\nu}$
does not solely determine the evolution.

\begin{acknowledgments}
V.G. and G.L.G. acknowledge  support by the European Research Council
under the grant ERC QGPDyn-259684.
JYO thanks Esteban Calzetta, Fran\c cois Gelis, Romuald Janik, Matthew Luzum and Li Yan for discussions. 
\end{acknowledgments}


\begin{thebibliography}{99}

\bibitem{Heinz:2013th} 
  U.~Heinz and R.~Snellings,
  Ann.\ Rev.\ Nucl.\ Part.\ Sci.\  {\bf 63}, 123 (2013)
  [arXiv:1301.2826 [nucl-th]].



\bibitem{Gale:2013da} 
  C.~Gale, S.~Jeon and B.~Schenke,
  Int.\ J.\ Mod.\ Phys.\ A {\bf 28}, 1340011 (2013)
  [arXiv:1301.5893 [nucl-th]].



\bibitem{Kolb:2003dz} 
  P.~F.~Kolb and U.~W.~Heinz,
  In *Hwa, R.C. (ed.) et al.: Quark gluon plasma* 634-714
  [nucl-th/0305084].



\bibitem{Adler:2003kt} 
  S.~S.~Adler {\it et al.}  [PHENIX Collaboration],
  Phys.\ Rev.\ Lett.\  {\bf 91}, 182301 (2003)
  [nucl-ex/0305013].



\bibitem{Aamodt:2010pa} 
  K.~Aamodt {\it et al.}  [ALICE Collaboration],
  Phys.\ Rev.\ Lett.\  {\bf 105}, 252302 (2010)
  [arXiv:1011.3914 [nucl-ex]].



\bibitem{Romatschke:2007mq} 
  P.~Romatschke and U.~Romatschke,
  Phys.\ Rev.\ Lett.\  {\bf 99}, 172301 (2007)
  [arXiv:0706.1522 [nucl-th]].



\bibitem{Luzum:2008cw} 
  M.~Luzum and P.~Romatschke,
  Phys.\ Rev.\ C {\bf 78}, 034915 (2008)
  [Erratum-ibid.\ C {\bf 79}, 039903 (2009)]
  [arXiv:0804.4015 [nucl-th]].



\bibitem{ALICE:2011ab} 
  K.~Aamodt {\it et al.}  [ALICE Collaboration],
  Phys.\ Rev.\ Lett.\  {\bf 107}, 032301 (2011)
  [arXiv:1105.3865 [nucl-ex]].



\bibitem{Adare:2011tg} 
  A.~Adare {\it et al.}  [PHENIX Collaboration],
  Phys.\ Rev.\ Lett.\  {\bf 107}, 252301 (2011)
  [arXiv:1105.3928 [nucl-ex]].



\bibitem{Alver:2010gr} 
  B.~Alver and G.~Roland,
  Phys.\ Rev.\ C {\bf 81}, 054905 (2010)
  [Erratum-ibid.\ C {\bf 82}, 039903 (2010)]
  [arXiv:1003.0194 [nucl-th]].



\bibitem{Alver:2010dn} 
  B.~H.~Alver, C.~Gombeaud, M.~Luzum and J.~Y.~Ollitrault,
  Phys.\ Rev.\ C {\bf 82}, 034913 (2010)
  [arXiv:1007.5469 [nucl-th]].



\bibitem{Baier:2007ix} 
  R.~Baier, P.~Romatschke, D.~T.~Son, A.~O.~Starinets and M.~A.~Stephanov,
  JHEP {\bf 0804}, 100 (2008)
  [arXiv:0712.2451 [hep-th]].



\bibitem{Teaney:2003kp} 
  D.~Teaney,
  Phys.\ Rev.\ C {\bf 68}, 034913 (2003)
  [nucl-th/0301099].



\bibitem{Teaney:2013gca} 
  D.~Teaney and L.~Yan,
  Phys.\ Rev.\ C {\bf 89}, no. 1, 014901 (2014)
  [arXiv:1304.3753 [nucl-th]].



\bibitem{Luzum:2010ad} 
  M.~Luzum and J.~Y.~Ollitrault,
  Phys.\ Rev.\ C {\bf 82}, 014906 (2010)
  [arXiv:1004.2023 [nucl-th]].



\bibitem{Dusling:2009df} 
  K.~Dusling, G.~D.~Moore and D.~Teaney,
  Phys.\ Rev.\ C {\bf 81}, 034907 (2010)
  [arXiv:0909.0754 [nucl-th]].



\bibitem{Borghini:2005kd} 
  N.~Borghini and J.~Y.~Ollitrault,
  Phys.\ Lett.\ B {\bf 642}, 227 (2006)
  [nucl-th/0506045].



\bibitem{Gombeaud:2007ub} 
  C.~Gombeaud and J.~Y.~Ollitrault,
  Phys.\ Rev.\ C {\bf 77}, 054904 (2008)
  [nucl-th/0702075].



\bibitem{Plumari:2012ep} 
  S.~Plumari, A.~Puglisi, F.~Scardina and V.~Greco,
  Phys.\ Rev.\ C {\bf 86}, 054902 (2012)
  [arXiv:1208.0481 [nucl-th]].



\bibitem{Monnai:2009ad} 
  A.~Monnai and T.~Hirano,
  Phys.\ Rev.\ C {\bf 80}, 054906 (2009)
  [arXiv:0903.4436 [nucl-th]].



\bibitem{Bozek:2009dw} 
  P.~Bozek,
  Phys.\ Rev.\ C {\bf 81}, 034909 (2010)
  [arXiv:0911.2397 [nucl-th]].



\bibitem{Noronha-Hostler:2013gga} 
  J.~Noronha-Hostler, G.~S.~Denicol, J.~Noronha, R.~P.~G.~Andrade and F.~Grassi,
  Phys.\ Rev.\ C {\bf 88}, 044916 (2013)
  [arXiv:1305.1981 [nucl-th]].



\bibitem{Ollitrault:2008zz} 
  J.~Y.~Ollitrault,
  Eur.\ J.\ Phys.\  {\bf 29}, 275 (2008)
  [arXiv:0708.2433 [nucl-th]].



\bibitem{Borsanyi:2013bia} 
  S.~Borsanyi, Z.~Fodor, C.~Hoelbling, S.~D.~Katz, S.~Krieg and K.~K.~Szabo,
  Phys.\ Lett.\ B {\bf 730}, 99 (2014)
  [arXiv:1309.5258 [hep-lat]].



\bibitem{Bjorken:1982qr} 
  J.~D.~Bjorken,
  Phys.\ Rev.\ D {\bf 27}, 140 (1983).



\bibitem{Heinz:2001xi} 
  U.~W.~Heinz and P.~F.~Kolb,
  Nucl.\ Phys.\ A {\bf 702}, 269 (2002)
  [hep-ph/0111075].



\bibitem{Miller:2007ri} 
  M.~L.~Miller, K.~Reygers, S.~J.~Sanders and P.~Steinberg,
  Ann.\ Rev.\ Nucl.\ Part.\ Sci.\  {\bf 57}, 205 (2007)
  [nucl-ex/0701025].



\bibitem{Retinskaya:2013gca} 
  E.~Retinskaya, M.~Luzum and J.~Y.~Ollitrault,
  Phys.\ Rev.\ C {\bf 89}, no. 1, 014902 (2014)
  [arXiv:1311.5339 [nucl-th]].



\bibitem{Teaney:2010vd} 
  D.~Teaney and L.~Yan,
  Phys.\ Rev.\ C {\bf 83}, 064904 (2011)
  [arXiv:1010.1876 [nucl-th]].



\bibitem{Bhalerao:2011yg} 
  R.~S.~Bhalerao, M.~Luzum and J.~Y.~Ollitrault,
  Phys.\ Rev.\ C {\bf 84}, 034910 (2011)
  [arXiv:1104.4740 [nucl-th]].



\bibitem{Niemi:2012aj} 
  H.~Niemi, G.~S.~Denicol, H.~Holopainen and P.~Huovinen,
  Phys.\ Rev.\ C {\bf 87}, no. 5, 054901 (2013)
  [arXiv:1212.1008 [nucl-th]].



\bibitem{Gardim:2014tya} 
  F.~G.~Gardim, J.~Noronha-Hostler, M.~Luzum and F.~Grassi,
  arXiv:1411.2574 [nucl-th].



\bibitem{Teaney:2012ke} 
  D.~Teaney and L.~Yan,
  Phys.\ Rev.\ C {\bf 86}, 044908 (2012)
  [arXiv:1206.1905 [nucl-th]].



\bibitem{Gardim:2011xv} 
  F.~G.~Gardim, F.~Grassi, M.~Luzum and J.~Y.~Ollitrault,
  Phys.\ Rev.\ C {\bf 85}, 024908 (2012)
  [arXiv:1111.6538 [nucl-th]].



\bibitem{Blaizot:2011xf} 
  J.~P.~Blaizot, F.~Gelis, J.~F.~Liao, L.~McLerran and R.~Venugopalan,
  Nucl.\ Phys.\ A {\bf 873}, 68 (2012)
  [arXiv:1107.5296 [hep-ph]].



\bibitem{Epelbaum:2014mfa} 
  T.~Epelbaum, F.~Gelis, N.~Tanji and B.~Wu,
  Phys.\ Rev.\ D {\bf 90}, no. 12, 125032 (2014)
  [arXiv:1409.0701 [hep-ph]].



\bibitem{Ferini:2008he} 
  G.~Ferini, M.~Colonna, M.~Di Toro and V.~Greco,
  Phys.\ Lett.\ B {\bf 670}, 325 (2009)
  [arXiv:0805.4814 [nucl-th]].



\bibitem{Greco:2008fs} 
  V.~Greco, M.~Colonna, M.~Di Toro and G.~Ferini,
  Prog.\ Part.\ Nucl.\ Phys.\  {\bf 62}, 562 (2009)
  [arXiv:0811.3170 [hep-ph]].



\bibitem{Plumari:2011re} 
  S.~Plumari and V.~Greco,
  AIP Conf.\ Proc.\  {\bf 1422}, 56 (2012)
  [arXiv:1110.2383 [hep-ph]].



\bibitem{Plumari:2012xz} 
  S.~Plumari, A.~Puglisi, M.~Colonna, F.~Scardina and V.~Greco,
  J.\ Phys.\ Conf.\ Ser.\  {\bf 420}, 012029 (2013)
  [arXiv:1209.0601 [hep-ph]].



\bibitem{Ruggieri:2013bda} 
  M.~Ruggieri, F.~Scardina, S.~Plumari and V.~Greco,
  Phys.\ Lett.\ B {\bf 727}, 177 (2013)
  [arXiv:1303.3178 [nucl-th]].



\bibitem{Xu:2004mz} 
  Z.~Xu and C.~Greiner,
  Phys.\ Rev.\ C {\bf 71}, 064901 (2005)
  [hep-ph/0406278].



\bibitem{Zhang:1999rs} 
  B.~Zhang, M.~Gyulassy and C.~M.~Ko,
  Phys.\ Lett.\ B {\bf 455}, 45 (1999)
  [nucl-th/9902016].



\bibitem{Molnar:2001ux} 
  D.~Molnar and M.~Gyulassy,
  Nucl.\ Phys.\ A {\bf 697}, 495 (2002)
  [Erratum-ibid.\ A {\bf 703}, 893 (2002)]
  [nucl-th/0104073].



\bibitem{Retiere:2003kf} 
  F.~Retiere and M.~A.~Lisa,
  Phys.\ Rev.\ C {\bf 70}, 044907 (2004)
  [nucl-th/0312024].



\bibitem{Yan:2014nsa} 
  L.~Yan, J.~Y.~Ollitrault and A.~M.~Poskanzer,
  Phys.\ Lett.\ B {\bf 742}, 290 (2015)
  [arXiv:1408.0921 [nucl-th]].


\bibitem{Chaudhuri:2005ea} 
  A.~K.~Chaudhuri and U.~W.~Heinz,
  J.\ Phys.\ Conf.\ Ser.\  {\bf 50}, 251 (2006)
  [nucl-th/0504022].


\bibitem{Florkowski:2010cf} 
  W.~Florkowski and R.~Ryblewski,
  Phys.\ Rev.\ C {\bf 83}, 034907 (2011)
  [arXiv:1007.0130 [nucl-th]].



\bibitem{Bhalerao:2013pza} 
  R.~S.~Bhalerao, A.~Jaiswal, S.~Pal and V.~Sreekanth,
  Phys.\ Rev.\ C {\bf 89}, no. 5, 054903 (2014)
  [arXiv:1312.1864 [nucl-th]].


\bibitem{Hatta:2014jva} 
  Y.~Hatta, J.~Noronha, G.~Torrieri and B.~W.~Xiao,
  Phys.\ Rev.\ D {\bf 90}, no. 7, 074026 (2014)
  [arXiv:1407.5952 [hep-ph]].


\bibitem{Gubser:2010ui} 
  S.~S.~Gubser and A.~Yarom,
  Nucl.\ Phys.\ B {\bf 846}, 469 (2011)
  [arXiv:1012.1314 [hep-th]].



\bibitem{Staig:2010pn} 
  P.~Staig and E.~Shuryak,
  Phys.\ Rev.\ C {\bf 84}, 034908 (2011)
  [arXiv:1008.3139 [nucl-th]].



\bibitem{Lacey:2013qua} 
  R.~A.~Lacey, A.~Taranenko, J.~Jia, D.~Reynolds, N.~N.~Ajitanand, J.~M.~Alexander, Y.~Gu and A.~Mwai,
  Phys.\ Rev.\ Lett.\  {\bf 112}, no. 8, 082302 (2014)
  [arXiv:1305.3341 [nucl-ex]].



\bibitem{Gyulassy:2003mc} 
  M.~Gyulassy, I.~Vitev, X.~N.~Wang and B.~W.~Zhang,
  In *Hwa, R.C. (ed.) et al.: Quark gluon plasma* 123-191
  [nucl-th/0302077].



\bibitem{Denicol:2011fa} 
  G.~S.~Denicol, J.~Noronha, H.~Niemi and D.~H.~Rischke,
  Phys.\ Rev.\ D {\bf 83}, 074019 (2011)
  [arXiv:1102.4780 [hep-th]].



\bibitem{Heller:2014wfa} 
  M.~P.~Heller, R.~A.~Janik, M.~Spaliński and P.~Witaszczyk,
  Phys.\ Rev.\ Lett.\  {\bf 113}, no. 26, 261601 (2014)
  [arXiv:1409.5087 [hep-th]].


\end{thebibliography}
\end{document}